\documentclass[aps,prd,onecolumn,superscriptaddress,preprintnumbers,floatfix,reprint]{revtex4}
\usepackage{graphicx}
\usepackage{epstopdf}
\usepackage{amsmath}
\usepackage{bm}
\usepackage{yhmath}
\usepackage[colorlinks]{hyperref}
\usepackage{subfigure}
\usepackage{color}
\usepackage{cases}
\usepackage{subfigure}
\usepackage{dcolumn,booktabs,bm}

\usepackage{amsfonts,amssymb,stmaryrd,latexsym,amsmath}
\usepackage{textcomp}

\usepackage{array}
\newcounter{RomanNumber}

\newcommand{\lyxmathsym}[1]{\ifmmode\begingroup\def\b@ld{bold}
  \text{\ifx\math@version\b@ld\bfseries\fi#1}\endgroup\else#1\fi}

\def\mud{{\lambda}}

\allowdisplaybreaks

\begin{document}

\title{Radiative decays of the doubly charmed baryons in chiral perturbation theory}

\author{Hao-Song Li}\email{haosongli@pku.edu.cn}\affiliation{School of Physics and State Key Laboratory of Nuclear Physics and Technology, Peking University, Beijing 100871, China}

\author{Lu Meng}\email{lmeng@pku.edu.cn}
\affiliation{School of Physics, Peking University, Beijing 100871,
China}

\author{Zhan-Wei Liu}\email{liuzhanwei@lzu.edu.cn}
\affiliation{School of Physical Science and Technology, Lanzhou
University, Lanzhou 730000, China}

\author{Shi-Lin Zhu}\email{zhusl@pku.edu.cn}\affiliation{School of Physics and State Key Laboratory of Nuclear Physics and Technology, Peking University, Beijing 100871, China}\affiliation{Collaborative Innovation Center of Quantum Matter, Beijing 100871, China}

\begin{abstract}

We have systematically investigated the spin-$\frac{3}{2}$ to
spin-$\frac{1}{2}$ doubly charmed baryon transition magnetic moments
to the next-to-next-to-leading order in the heavy baryon chiral
perturbation theory (HBChPT). Numerical results of transition
magnetic moments and decay widths are presented to the
next-to-leading order:
$\mu_{\Xi_{cc}^{*++}\rightarrow\Xi_{cc}^{++}}=-2.35\mu_{N}$,
$\mu_{\Xi_{cc}^{*+}\rightarrow\Xi_{cc}^{+}}=1.55\mu_{N}$,
$\mu_{\Omega_{cc}^{*+}\rightarrow\Omega_{cc}^{+}}=1.54\mu_{N}$,
$\Gamma_{\Xi_{cc}^{*++}\rightarrow\Xi_{cc}^{++}}=22.0$ keV,
$\Gamma_{\Xi_{cc}^{*+}\rightarrow\Xi_{cc}^{+}}=9.57$ keV,
$\Gamma_{\Omega_{cc}^{*+}\rightarrow\Omega_{cc}^{+}}=9.45$ keV.

\end{abstract}

\maketitle

\thispagestyle{empty}

\section{Introduction}\label{Sec1}

SELEX Collaboration first reported the possible candidates of the
doubly charm baryons~\cite{Mattson:2002vu}, which were unfortunately
not confirmed by other experimental collaborations like
FOCUS~\cite{Ratti:2003ez}, BABAR~\cite{Aubert:2006qw} and
Belle~\cite{Chistov:2006zj}. Recently, LHCb collaboration discovered
$\Xi^{++}_{cc}$ in the mass spectrum of
$\Lambda_{c}^{+}K^{-}\pi^{+}\pi^{+}$ with the mass
$M_{\Xi^{++}_{cc}}=3621.40\pm0.72\pm0.27\pm0.14
\rm{MeV}$~\cite{LHCb}. The mass and decay properties of double heavy
charm and bottom baryons have been studied extensively in literature
\cite{Bagan:1992za,Roncaglia:1995az,SilvestreBrac:1996bg,Ebert:1996ec,Tong:1999qs,Itoh:2000um,Gershtein:2000nx,
Kiselev:2001fw,Kiselev:2002iy,Narodetskii:2001bq,Lewis:2001iz,Faessler:2001mr,Ebert:2002ig,Mathur:2002ce,Flynn:2003vz,Vijande:2004at,Chiu:2005zc,
Migura:2006ep,Albertus:2006ya,Liu:2007fg,Roberts:2007ni,Valcarce:2008dr,Liu:2009jc,Namekawa:2012mp,Alexandrou:2012xk,
Aliev:2012ru,Aliev:2012iv,Namekawa:2013vu,Dhir:2013nka,Sun:2014aya,Chen:2015kpa,Sun:2016wzh,
Shah:2016vmd,Chen:2016spr,Kiselev:2017eic,Chen:2017sbg,Gutsche:2017hux,Wang:2017mqp,Meng:2017udf,Yu:2017zst,Wang:2017azm,
Karliner:2017qjm,Eichten:2017ffp,Hu:2005gf}.

As the electromagnetic properties characterize fundamental aspects
of the inner structure of baryons, it is very important to
investigate the baryon electromagnetic form factors, especially the
magnetic moments. Lichtenberg first investigated the magnetic
moments of spin-$\frac{1}{2}$ doubly charmed baryons with
nonrelativistic qurak model in Ref. \cite{Lichtenberg:1976fi}. Since
then, more and more approaches were developed to investigate the
magnetic moments of spin-$\frac{1}{2}$ doubly charmed baryons
\cite{JuliaDiaz:2004vh,Faessler:2006ft,Bose:1980vy,Bernotas:2012nz,Jena:1986xs,
Oh:1991ws,Patel:2008xs,Can:2013zpa,Can:2013tna,Li3}. However, as the
degenerate partner state of the spin-$\frac{1}{2}$ doubly charmed
baryons in the heavy quark limit, the spin-$\frac{3}{2}$ doubly
charmed baryons were rarely studied. The spin-$\frac{3}{2}$ to
spin-$\frac{1}{2}$ doubly charmed baryon transition magnetic moment
deserves more attention as it probes the inner structure and
possible deformation of both the spin-$\frac{1}{2}$ and
spin-$\frac{3}{2}$ doubly charmed baryons.

The spin-$\frac{3}{2}$ to spin-$\frac{1}{2}$ baryon transition
amplitude has been systematically investigated in Refs.
\cite{Bjorken:1966ij,Jones:1972ky}. The transition amplitude
contains the magnetic dipole (M1), electric quadrupole (E2), and
Coulumb quadrupole (C2) contributions with the spin-parity selection
rule. For the doubly charmed baryons, the radiative transitions have
been studied in the MIT bag model
\cite{Hackman:1977am,Bernotas:2013eia}, SU(4) chiral constituent
quark model ($\chi$CQM) \cite{Sharma:2010vv} and manifestly Lorentz
covariant constituent three quark model \cite{Branz:2010pq}.

In this paper, we focus on the transition magnetic moment of the
doubly charmed baryons in the framework of chiral perturbation
theory (ChPT) ~\cite{Weinberg:1978kz}. ChPT is a very useful
framework in the low-energy hadron physics and was first developed
to deal with the pseudoscalar meson system. When extended to include
the baryons, heavy baryon chiral perturbation theory (HBChPT) was
proposed
~\cite{Jenkins:1990jv,Jenkins:1992pi,Bernard:1992qa,Bernard:1995dp}.

In this work, we will employ HBChPT to calculate the the one-loop
chiral corrections to the spin-$\frac{3}{2}$ to spin-$\frac{1}{2}$
doubly charmed baryon transition magnetic moments order by order. We
use quark model to estimate some low energy constants (LEC) in ChPT
including the leading-order axial coupling and tree level transition
magnetic moments. We explicitly consider the mass splitting $\delta$
between spin-$\frac{3}{2}$ and spin-$\frac{1}{2}$ doubly charmed
baryons. We present our numerical results up to the next-to-leading
order while we derive the analytical results to the
next-to-next-to-leading order.

This paper is organized as follows. In Section \ref{Sec3}, we
discuss the spin-$\frac{3}{2}$ to spin-$\frac{1}{2}$ baryon
electromagnetic transition form factors. We introduce the effective
chiral Lagrangians of the doubly charmed baryons in Section
\ref{Sec2}. In Section \ref{secFormalism}, we calculate the
transition magnetic moments order by order. We estimate the
low-energy constants and present our numerical results in Section
\ref{Sec6}. We give a short summary in Section \ref{Sec7} and
collect some useful formulae and the coefficients of the loop
corrections in the Appendix~\ref{appendix-A}.

\section{Spin-$\frac{3}{2}$ to spin-$\frac{1}{2}$ baryon electromagnetic transition form factors}\label{Sec3}

With the constraint of  Lorentz covariance, gauge invariance, parity
conservation and time reversal invariance, the matrix element of
electromagnetic current between spin-$\frac{3}{2}$ and
spin-$\frac{1}{2}$ baryon states can be written
as~\cite{Jones:1972ky,Li2}:
\begin{equation}
<B(p)|J_{\mu}|T(p^{\prime})>=e\bar{u}(p)O_{\rho\mu}(p^{\prime},p)u^{\rho}(p^{\prime}),
\end{equation}
with
\begin{equation}
O_{\rho\mu}(p^{\prime},p)=\frac{G_{1}}{2M_B}(q_{\rho}\gamma_{\mu}-q\cdot\gamma
g_{\rho\mu})\gamma_{5}+\frac{G_{2}}{4M_{B}^{2}}\frac{1}{M_{B}+M_{T}}(q\cdot
Pg_{\rho\mu}-q_{\rho}P_{\mu})q\hspace{-0.5em}/\gamma_{5}.
\label{eq_new_current}
\end{equation}
Here we use $B$ to denote the spin-$\frac{1}{2}$ baryons and $T$ to
denote spin-$\frac{3}{2}$ baryons. $p$ and $p'$ are the
corresponding momenta of the baryons. In the above equations,
$P=\frac{1}{2}(p^{\prime}+p)$, $q=p^{\prime}-p$, $M_{B}$ and $M_{T}$
are the corresponding baryon masses, and $u_{\rho}(p)$ is the
Rarita-Schwinger spinor satisfying $p^{\rho}u_{\rho}(p)=0$ and
$\gamma^{\rho}u_{\rho}(p)=0$ for an on-shell heavy baryon.

In the framework of HBChPT, the baryon field $B$ can be decomposed
into the large component $\mathcal{N}$ and the small component
$\mathcal{H}$.
\begin{equation}
B=e^{-iM_{B}v\cdot x}(\mathcal{N}+\mathcal{H}),
\end{equation}
\begin{equation}
\mathcal{N}=e^{iM_{B}v\cdot x}\frac{1+v\hspace{-0.5em}/}{2}B,~
\mathcal{H}=e^{iM_{B}v\cdot x}\frac{1-v\hspace{-0.5em}/}{2}B,
\end{equation}
where $v_{\mu}=(1,\vec{0})$ is the velocity of the baryon. For the
Rarita-Schwinger field, the large component of spin-$\frac{3}{2}$
degrees of freedom is denoted as $T^{\rho}$. Now the
spin-$\frac{3}{2}$ and spin-$\frac{1}{2}$ matrix elements of the
electromagnetic current $J_{\mu}$ can be parameterized as
\begin{equation}
<\mathcal N(p)|J_{\mu}|T^\rho(p^{\prime})>=e\bar{u}(p)\mathcal
O_{\rho\mu}(p^{\prime},p)u^{\rho}(p^{\prime}).
\end{equation}
The tensor $\mathcal O_{\rho\mu}$ can be parameterized in terms of
two Lorentz invariant form factors.
\begin{eqnarray}
\begin{split}
\mathcal
O_{\rho\mu}(p^{\prime},p)=\frac{G_{1}}{M_{B}}(q_{\rho}S_{\mu}-q\cdot
Sg_{\rho\mu})+\frac{G_{2}}{4M_{B}^{2}}(q\cdot
vg_{\rho\mu}-q_{\rho}v_{\mu})q\cdot S. \label{eq_newnew_current}
\end{split}
\end{eqnarray}
In this paper, we will use Eq.~(\ref{eq_newnew_current}) to define
the electro quadrupole (E2) and magnetic-dipole (M1) multipole
transtion form factors. The multipole form factors are
\begin{eqnarray}
G_{M1}  &=&  \frac{2}{3}G_{1}-\frac{\delta}{6M_{T}}G_{1}-\frac{\delta}{12M_{B}}G_{2},\label{eq_formfactor1}\\
G_{E2} & = & \frac{\delta}{6M_{T}}G_{1}-\frac{\delta}{12M_{B}}G_{2},
\label{eq_formfactor2}
\end{eqnarray}
where $|q|=\delta$ in the rest frame of spin-$\frac{3}{2}$ baryon.
The decay width and transition magnetic moment are expressed as
\begin{eqnarray}
\Gamma(T\rightarrow B\gamma) & = & \frac{\alpha}{16}\frac{(M_{T}^2-M_{B}^2)^3}{M_{T}^3M_{B}^2}(|G_{M1}(q^{2}=0)|^{2}+3|G_{E2}(q^{2}=0)|^{2}),\label{eq:decaywidth}\\
\mu(T\rightarrow
B\gamma)&=&\frac{2M_{T}}{M_{T}+M_{B}}G_{M1}(q^{2}=0)\frac{e}{2M_{B}}.
\end{eqnarray}
where $\alpha=\frac{e^2}{4\pi}=\frac{1}{137}$ is the electromagnetic
fine structure constant.

\section{Chiral Lagrangians}\label{Sec2}

\subsection{The strong interaction chiral Lagrangians}

We follow the basic definitions of the pseudoscalar mesons and the
spin-$\frac{1}{2}$ doubly charmed baryon chiral effective
Lagrangians in Refs.~\cite{Bernard:1995dp,Li3}. The pseudoscalar
meson fields are defined as:
\begin{equation}
\phi=\left(\begin{array}{ccc}
\pi^{0}+\frac{1}{\sqrt{3}}\eta & \sqrt{2}\pi^{+} & \sqrt{2}K^{+}\\
\sqrt{2}\pi^{-} & -\pi^{0}+\frac{1}{\sqrt{3}}\eta & \sqrt{2}K^{0}\\
\sqrt{2}K^{-} & \sqrt{2}\bar{K}^{0} & -\frac{2}{\sqrt{3}}\eta
\end{array}\right).
\end{equation}
 The chiral connection and axial vector
field are defined as~\cite{Bernard:1995dp}:
\begin{equation}
\Gamma_{\mu}=\frac{1}{2}\left[u^{\dagger}(\partial_{\mu}-ir_{\mu})u+u(\partial_{\mu}-il_{\mu})u^{\dagger}\right],
\end{equation}
\begin{equation}
u_{\mu}\equiv\frac{1}{2}i\left[u^{\dagger}(\partial_{\mu}-ir_{\mu})u-u(\partial_{\mu}-il_{\mu})u^{\dagger}\right],
\end{equation}
where
\begin{equation}
u^{2}=\mathit{U}=\exp(i\phi/f_{0}).
\end{equation}
We use the pseudoscalar meson decay constants $f_{\pi}\approx$ 92.4
MeV, $f_{K}\approx$ 113 MeV and $f_{\eta}\approx$ 116 MeV.

For the spin-$\frac{3}{2}$ doubly charmed baryon field, we adopt the
Rarita-Schwinger field~\cite{Rarita:1941mf}.
\begin{equation}
\Psi^{*\mu}=\left(\begin{array}{c}
\Xi_{cc}^{*++}\\
\Xi_{cc}^{*+}\\
\Omega_{cc}^{*+}
\end{array}\right)^\mu\Rightarrow\left(\begin{array}{c}
ccu\\
ccd\\
ccs
\end{array}\right)^\mu.
\end{equation}
The leading order pseudoscalar meson and baryon interaction
Lagrangians read
\begin{eqnarray}
\hat{\mathcal{L}}_{0}^{(1)}&=&\bar{\Psi}(iD\hspace{-0.6em}/-M_{H})\Psi \nonumber\\
&&+{\bar
\Psi^{*\mu}[-g_{\mu\nu}(iD\hspace{-0.6em}/-M_{T})+i(\gamma_{\mu}
D_{\mu}+\gamma_{\nu}D_{\mu})-\gamma_{\mu}(iD\hspace{-0.6em}/+M_{T})\gamma_{\nu}]\Psi^{*\nu}},
\label{Eq:baryon01}\\
\hat{\mathcal{L}}_{\rm
int}^{(1)}&=&\frac{\tilde{g}_{A}}{2}\bar{\Psi}u\hspace{-0.5em}/\gamma_{5}\Psi+\frac{\tilde{g}_{B}}{2}
\bar \Psi^{*\mu}g_{\mu\nu}u\hspace{-0.5em}/
\gamma_{5}\Psi^{*\nu}+\frac{\tilde{g}_{C}}{2}[\bar
\Psi^{*\mu}u_{\mu}\Psi+\bar \Psi u_{\mu}\Psi^{*\mu}]
,\label{Eq:baryon02}
\end{eqnarray}
where $M_{H}$ is the spin-$\frac{1}{2}$ doubly charmed baryon mass,
$M_{T}$ is the spin-$\frac{3}{2}$ doubly charmed baryon mass,
\begin{eqnarray}
D_{\mu}\Psi&=&\partial_{\mu}\Psi+\Gamma_{\mu}\Psi, \nonumber\\
D^{\nu}\Psi^{*\mu}&=&\partial^{\nu}\Psi^{*\mu}+\Gamma^{\nu}\Psi^{*\mu}.
\end{eqnarray}
We also need the second order pseudoscalar meson and doubly charmed
baryon interaction Lagrangians. Recall that
\begin{eqnarray}
3\otimes\bar{3} & = & 1\oplus8\label{Eq:flavor1},\\
8\otimes8 & = &
1\oplus8_{1}\oplus8_{2}\oplus10\oplus\bar{10}\oplus27.\label{Eq:flavor2}
\end{eqnarray}
When the product of $u_{\mu}$ and $u_{\nu}$ belongs to the $8_1$ and
$8_2$ flavor representation, we can write down two independent
interaction terms of the second order pseudoscalar meson and doubly
charmed baryon Lagrangians:
\begin{eqnarray}
\hat{\mathcal{L}}_{\rm int}^{(2)}&=&\frac{h_{1}}{4M_{B}}
\bar{\Psi}[u_{\mu}, u_{\nu}]
\gamma^{\nu}\gamma_{5}\Psi^{*\mu}+\frac{h_{2}}{4M_{B}}
\bar{\Psi}\{u_{\mu}, u_{\nu}\}
\gamma^{\nu}\gamma_{5}\Psi^{*\mu}+{\rm H.c.}, \label{Eq:baryon03}
\end{eqnarray}
where $M_{B}$ is the nucleon mass and $h_{1,2}$ are the coupling
constants.

In the framework of HBChPT, we denote the large component of the
spin-$\frac{3}{2}$ doubly charmed baryon as $T_{\mu}$. The leading
order nonrelativistic pseudoscalar meson and baryon Lagrangians read
\begin{equation}
\mathcal{L}_{0}^{(1)}=\bar{H}(iv\cdot D)H-i\bar{T}^{\mu}(v\cdot
D-\delta)T_{\mu}, \label{Eq:baryon1}
\end{equation}
\begin{equation}
\mathcal{L}_{\rm int}^{(1)}=\tilde{g}_{A}\bar{H}S_\mu u^\mu
H+\tilde{g}_{B} \bar{T}^{\rho}S_\mu u^\mu
T_{\rho}+\frac{\tilde{g}_{C}}{2}[\bar{T}^{\mu}u_{\mu}H+\bar H
u_{\mu}T^{\mu}], \label{Eq:baryon2}
\end{equation}
where $S_{\mu}$ is the covariant spin-operator, $\delta=M_{T}-M_{H}$
is the spin-$\frac{1}{2}$ and spin-$\frac{3}{2}$ doubly charmed
baryon mass splitting. We adopt $\delta = 0.1$ GeV approximatively
\cite{Albertus:2006ya}. The $\phi H H$ coupling
$\tilde{g}_{A}=-\frac{2}{5}g_{A}=-0.50$~\cite{Li3}. With the help of
quark model, we have estimated the $\phi T T$ coupling
$\tilde{g}_{B}=-\frac{6}{5}g_{A}=-1.51$ and $\phi T H$ coupling
$\tilde{g}_{C}=-\frac{4\sqrt{3}}{5}g_{A}=-1.75$. For the
pseudoscalar mesons  masses, we use $m_{\pi}=0.140$ GeV,
$m_{K}=0.494$ GeV, and $m_{\eta}=0.550$ GeV. We use the nucleon
masses $M_B=0.938$ GeV and the spin-$\frac{1}{2}$ doubly charmed
baryon mass $M_{H}=3.62$ GeV.

The second order pseudoscalar meson and baryon nonrelativistic
Lagrangians read
\begin{eqnarray}
\hat{\mathcal{L}}_{\rm int}^{(2)}&=&\frac{h_{1}}{2M_{B}}
\bar{H}[u_{\mu}, u_{\nu}] S^{\nu}T^{\mu}+\frac{h_{2}}{2M_{B}}
\bar{H}\{u_{\mu}, u_{\nu}\} S^{\nu}T^{\mu}+{\rm H.c.}.
\label{Eq:TNUU2}
\end{eqnarray}
The above Lagrangians contribute to the spin-$\frac{3}{2}$ to
spin-$\frac{1}{2}$ doubly charmed baryon transition magnetic moments
in diagram (d) of Fig.~\ref{fig:allloop}. After loop integration,
the contribution of the $h_{2}$ term vanishes. Thus, there are only
one linearly independent low energy constant (LEC)
 $h_{1}$  which contribute
to the present investigations of transition magnetic moments up to
$\mathcal{O}(p^4)$.

\subsection{The electromagnetic chiral Lagrangians at $\mathcal{O}(p^{2})$}

Following the definitions in Ref. \cite{Li3}, the lowest order
$\mathcal{O}(p^{2})$ Lagrangians contribute to the magnetic moments
of the spin-$\frac{1}{2}$ doubly charmed baryons at the tree level
read
\begin{equation}
\mathcal{L}_{\mu_{H}}^{(2)}=a_{1}\frac{-i}{4M_{B}}\bar{H}[S^{\mu},S^{\nu}]\hat{F}_{\mu\nu}^{+}H
+a_{2}\frac{-i}{4M_{B}}\bar{H}[S^{\mu},S^{\nu}]H{\rm
Tr}(F_{\mu\nu}^{+}), \label{Eq:MM12}
\end{equation}
where the operator
$\hat{F}_{\mu\nu}^{+}=F_{\mu\nu}^{+}-\frac{1}{3}\rm
Tr(F_{\mu\nu}^{+})$ is traceless and transforms as the adjoint
representation. We can also write the lowest order
$\mathcal{O}(p^{2})$ Lagrangians which contribute to the magnetic
moments of the spin-$\frac{3}{2}$ doubly charmed baryons and the
spin-$\frac{3}{2}$ to spin-$\frac{1}{2}$ transition magnetic
moments,
\begin{equation}
\mathcal{L}_{\mu_{T}}^{(2)}=\frac{-ia_{3}}{2M_{B}}\bar{T}^{\mu}\hat{F}_{\mu\nu}^{+}
T^{\nu}+\frac{-ia_{4}}{2M_{B}}\bar{T}^{\mu}T^{\nu}{\rm
Tr}(F_{\mu\nu}^{+}), \label{Eq:MM32}
\end{equation}
\begin{equation}
\mathcal{L}_{\mu_{TH}^{(2)}}=a_5\frac{-i}{2M_{B}}\bar{T}^{\mu}
\hat{F}_{\mu\nu}^{+}S^{\nu}H
+a_6\frac{-i}{2M_{B}}\bar{T}^{\mu}S^{\nu}H{\rm Tr}(F_{\mu\nu}^{+})+
{\rm H.c.},\label{Eq:baryon_trans}
\end{equation}

\subsection{The higher order electromagnetic chiral Lagrangians }

We also need the $\mathcal{O}(p^{4})$ electromagnetic chiral
Lagrangians at the tree level to calculate the magnetic moments to
$\mathcal{O}(p^{3})$. Recalling flavor representation in
Eqs.~(\ref{Eq:flavor1}), (\ref{Eq:flavor2}) and considering that we
only need the leading-order terms of the fields $F_{\mu\nu}^{+}$ and
$\chi^{+}$ which are diagonal matrices, only three independent terms
contribute to the magnetic moments of the doubly charmed baryons up
to $\mathcal{O}(p^{3})$,
\begin{eqnarray}
\mathcal{L}_{\mu_{TH}}^{(4)}&=&d_{1}\frac{-i}{2M_{B}}\bar{T}^\mu
S^{\nu}H{\rm
Tr}(\chi^{+}\hat{F}_{\mu\nu}^{+})+d_{2}\frac{-i}{2M_{B}}\bar{T}^\mu
S^{\nu}\{\hat{F}_{\mu\nu}^{+},\chi^{+}\}H\nonumber
\\&&+d_{3}\frac{-i}{2M_{B}}\bar{T}^\mu S^{\nu}\chi^{+}H{\rm Tr}(F_{\mu\nu}^{+})\label{Eq:MM3}
\end{eqnarray}
where $\chi^{+}$=diag(0,0,1) at the leading order and the factor
$m_{s}$ has been absorbed in the LECs $d_{1,2,3}$. There are two
more terms which also contribute to the doubly charmed baryon
magnetic moments.
\begin{eqnarray}
\mathcal{L^{\prime}}_{\mu_{TH}}^{(4)}&=&a_{5}^{\prime}\frac{-i}{2M_{B}}\bar{T}^\mu
S^{\nu}\hat{F}_{\mu\nu}^{+}H{\rm
Tr}(\chi^{+})+a_{6}^{\prime}\frac{-i}{2M_{B}}\bar{T}^\mu
S^{\nu}H{\rm Tr}(F_{\mu\nu}^{+}){\rm Tr}(\chi^{+})
\end{eqnarray}
However, their contributions can be absorbed through the
renomalization of the LECs $a_{5,6}$, i.e.
\begin{eqnarray}
a_{5}&\rightarrow&a_{5}+{\rm Tr}(\chi^{+})a_{5}^{\prime},\\
a_{6}&\rightarrow&a_{6}+{\rm Tr}(\chi^{+})a_{6}^{\prime}.
\end{eqnarray}

\section{Formalism up to one-loop level}\label{secFormalism}

\begin{figure}
\centering
\includegraphics[width=0.6\hsize]{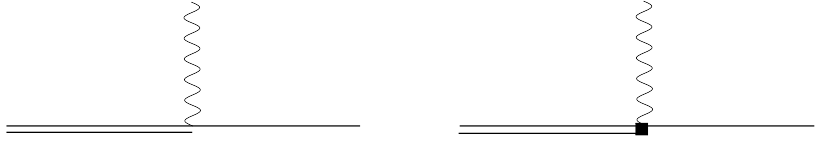}
\caption{The $\mathcal{O}(p^{2})$ and $\mathcal{O}(p^{4})$ tree
level diagram where the spin-$\frac{1}{2}$ (spin-$\frac{3}{2}$)
doubly charmed baryon is denoted by the single (double) solid line.
The left dot and the right black square represent second- and
fourth-order couplings respectively.} \label{fig:tree}
\end{figure}

Considering the standard power counting scheme of HBChPT, the chiral
order $D_{\chi}$ of a given diagram ~\cite{Ecker:1994gg}
\begin{equation}
D_{\chi}=4N_{L}-2I_{M}-I_{B}+\sum_{n}nN_{n}, \label{Eq:Power
counting}
\end{equation}
where $N_{L}$ is the number of loops, $I_{M}$ is the number of
internal meson lines, $I_{B}$ is the number of internal baryon lines
and $N_{n}$ is the number of the vertices from the $n$th order
Lagrangians. Thus, the chiral order of the transition magnetic
moments is $(D_\chi-1)$. The tree-level Lagrangians in
Eqs.~(\ref{Eq:baryon_trans}),(\ref{Eq:MM3}) contribute to the
transition magnetic moments at $\mathcal{O}(p^{1})$ and
$\mathcal{O}(p^{3})$ as shown in Fig.~\ref{fig:tree}. The
Clebsch-Gordan coefficients for the various processes are collected
in Table~\ref{Magnetic moments}. All tree level transition magnetic
moments are given in terms of $a_5$, $a_6$, $d_{1}$, $d_{2}$ and
$d_{3}$.

\begin{figure}[tbh]
\centering
\includegraphics[width=0.9\hsize]{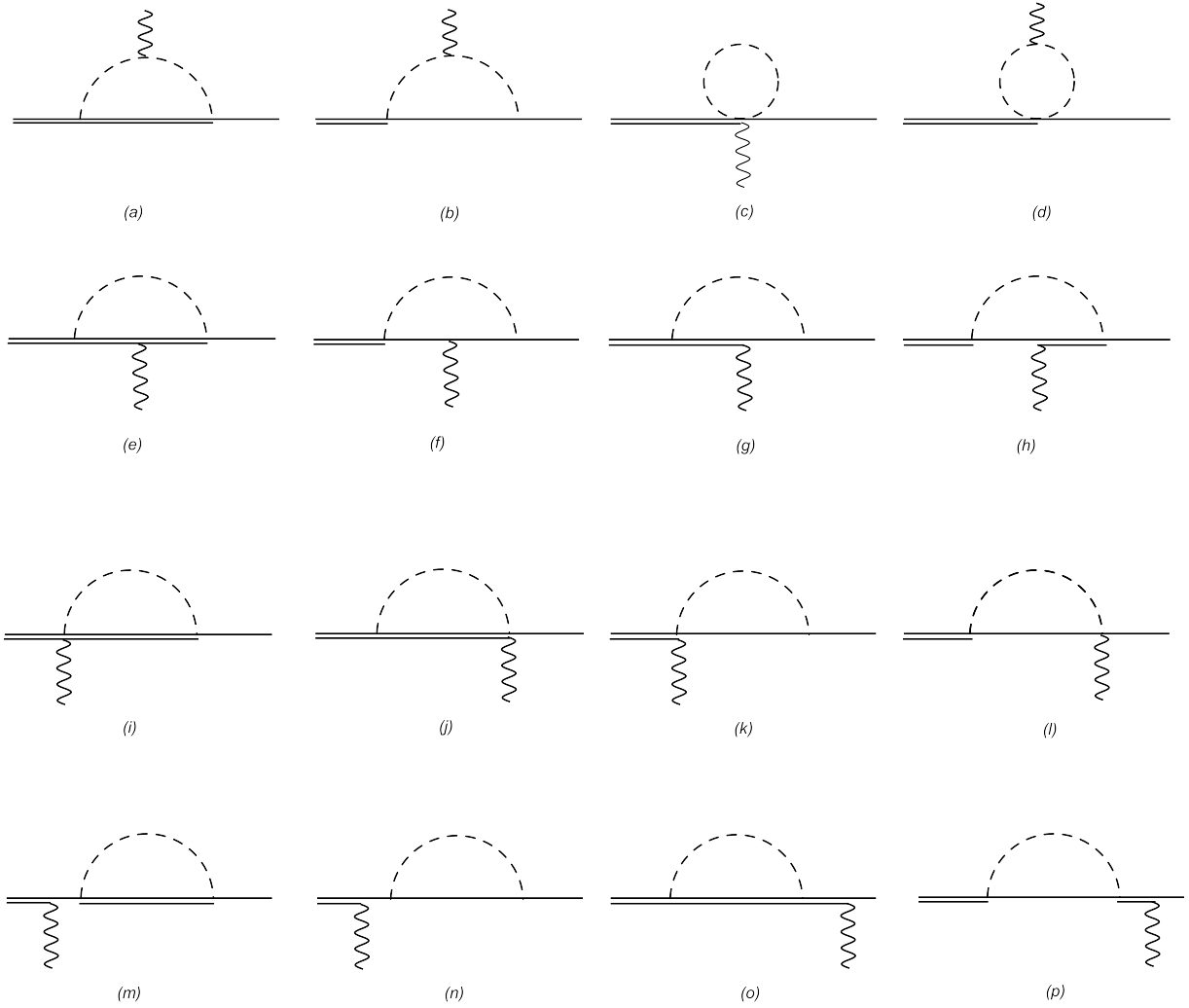}
\caption{The one-loop diagrams where the spin-$\frac{1}{2}$
(spin-$\frac{3}{2}$) doubly charmed baryon is denoted by the single
(double) solid line. The dashed and wiggly lines represent the
pseudoscalar meson and photon respectively.}\label{fig:allloop}

\end{figure}

There are sixteen Feynman diagrams contribute to the
spin-$\frac{3}{2}$ to spin-$\frac{1}{2}$ doubly charmed baryon
transition magnetic moments at one-loop level as shown in
Fig.~\ref{fig:allloop}. In diagrams (a-b), the photon vertex is from
the meson photon interaction term while the meson vertex is from the
interaction terms in Eq.~(\ref{Eq:baryon2}). In diagram (c), the
photon-meson-baryon vertex is from the high-order expansion of the
$\mathcal{O}(p^{2})$ tree level transition magnetic moment
interaction in Eq.~(\ref{Eq:baryon_trans}). In diagram (d), the
meson-baryon vertex is from the second order pseudoscalar meson and
baryon Lagrangian in Eq.~(\ref{Eq:TNUU2}) while the photon vertex is
also from the meson photon interaction term. In diagrams (e-h), the
photon-baryon vertex from the $\mathcal{O}(p^{2})$ tree level
magnetic moment interaction in Eqs.~(\ref{Eq:MM12}),
(\ref{Eq:MM32}), (\ref{Eq:baryon_trans}) while the meson vertex is
from the interaction terms in Eq.~(\ref{Eq:baryon2}). In diagrams
(i-l), the two vertices are from the strong interaction and seagull
terms respectively. In diagrams (m-p), the meson vertex is from the
strong interaction terms while the photon vertex from the
$\mathcal{O}(p^{2})$ tree level spin-$\frac{3}{2}$ to
spin-$\frac{1}{2}$ doubly charmed baryon transition magnetic moment
interaction in Eq.~(\ref{Eq:baryon_trans}).

The diagrams (a) and (b) contribute to the transition magnetic
moment at $\mathcal{O}(p^{2})$ while the diagrams (c-p) contributes
at $\mathcal{O}(p^{3})$. The diagrams (i-l) vanish in the heavy
baryon mass limit for the same reason as in Ref.~\cite{Li2}. The
diagrams (m-p) are the wave function renormalization corrections.

Summing all the one-loop level contributions in
Fig.~\ref{fig:allloop}, the loop corrections to the
spin-$\frac{3}{2}$ to spin-$\frac{1}{2}$ doubly charmed baryon
transition magnetic moments can be expressed as
\begin{eqnarray}
\mu_{TH}^{(2,\rm loop)}& =
&\frac{2M_{T}}{M_{T}+M_{H}}\frac{e}{2M_B}(\frac{2}{3}-\frac{\delta}{6M_{T}})M_{B}\sum_{\phi=\pi,K}
\left(\frac{-\beta_{a}^\phi
\tilde{g}_{B}\tilde{g}_{C}}{4f_\phi^{2}}a^{\phi}_{T}+\frac{\beta_{b}^\phi\tilde{g}_{A}
\tilde{g}_{C}}{4f_\phi^{2}}b^{\phi}_{H}\right),\label{eq:muTN2Loop}\\
\mu_{TH}^{(3,\rm loop)}& =
&\frac{2M_{T}}{M_{T}+M_{H}}\frac{e}{2M_B}(\frac{2}{3}-\frac{\delta}{6M_{T}})
[\sum_{\phi=\pi,K}(\gamma_{c}^\phi+\gamma_{d}^\phi)\frac{m_\phi^{2}}{64\pi^{2}f_\phi^{2}}
\ln\frac{m_\phi}{\mud} \nonumber\\&&
+\sum_{\phi=\pi,K,\eta}(\frac{5\tilde{g}_B\tilde{g}_C}{36\delta
f_{\phi}^{2}}e^{\phi}_{T}\gamma_{e}^\phi
-\frac{\tilde{g}_A\tilde{g}_C}{8f_\phi^{2}\delta}f^{\phi}_{H}\gamma_{f}^\phi+\frac{5\tilde{g}_A\tilde{g}_B}{16f_\phi^{2}}g^{\phi}_{TH}
\gamma^\phi_{TH}-\frac{\tilde{g}_A\tilde{g}_B}{32\delta
f_\phi^{2}}h^{\phi}_{HT} \gamma^\phi_{TH})] \nonumber\\&  &
+\frac{2M_{T}}{M_{T}+M_{H}}(\frac{2}{3}-\frac{\delta}{6M_{T}}){e\over 2M_B}\sum_{\phi=\pi,K,\eta}[\frac{\tilde{g}_A^2}
{f_\phi^{2}}\frac{-3}{32}n^{\phi}_{H}\gamma_{w}^\phi
 -\frac{\tilde{g}_C^2}{16f_\phi^{2}}m^{\phi}_{H}\gamma_{w}^\phi
 \nonumber\\
&&+\frac{\tilde{g}_B^2}{f_\phi^{2}}\frac{-3}{32}o^{\phi}_{H}\gamma_{w}^\phi-\frac{\tilde{g}_C^2}
{32f_\phi^{2}}p^{\phi}_{T}\gamma_{w}^\phi],\quad
\label{eq:muTN3Loop}
\end{eqnarray}

\begin{eqnarray}
a^{\phi}_{T} & = &\frac{1}{3456 \pi ^2 \delta^2 }[176 \delta ^3-84 \delta ^3 \ln \left(\frac{m_{\phi}^2}{\lambda ^2}\right)+24(2m_{\phi}^2+7\delta^2) \sqrt{m_{\phi}^2-\delta ^2} \arccos\left(\frac{\delta }{m_{\phi}}\right)\nonumber\\
&&-108\delta m_{\phi}^2 \arccos\left(\frac{\delta
}{m_{\phi}}\right)^2
+(27 \pi ^2 +48)\delta m_{\phi}^2-24m_{\phi}^3\pi],  \\
b^{\phi}_{H} & = &\frac{1}{1152 \pi ^2 \delta ^2} [
32 \delta ^3+24 \pi  m_{\phi}^3-12 \delta ^3 \ln \left(\frac{m_{\phi}^2}{\lambda ^2}\right)-24 \sqrt{m_{\phi}^2-\delta ^2} \left(\delta ^2+2 m_{\phi}^2\right) \arccos\left(-\frac{\delta }{m_{\phi}}\right)\nonumber\\
&&+9 \pi ^2 \delta  m_{\phi}^2+48 \delta  m_{\phi}^2-36 \delta
m_{\phi}^2 \arccos\left(-\frac{\delta }{m_{\phi}}\right)^2,\\
e^{\phi}_{T} & = &\frac{1}{144 \pi ^2}[\left(6 \delta ^3-9 \delta
m_{\phi}^2\right) \ln \left(\frac{m_{\phi}^2}{\lambda ^2}\right)+12
\left(m_{\phi}^2-\delta ^2\right)^{3/2} \arccos\left(\frac{\delta
}{m_{\phi}}\right)-2 \left(5 \delta ^3+3
\pi  m_{\phi}^3-6 \delta  m_{\phi}^2\right)],\\
f^{\phi}_{H} & = &\frac{1}{144 \pi ^2}[\left(6 \delta ^3-9 \delta
m_{\phi}^2\right) \ln \left(\frac{m_{\phi}^2}{\lambda ^2}\right)-12
\left(m_{\phi}^2-\delta ^2\right)^{3/2} \arccos\left(-\frac{\delta
}{m_{\phi}}\right)+2 \left(-5 \delta ^3+3 \pi  m_{\phi}^3+6 \delta
m_{\phi}^2\right)],\\
g^{\phi}_{TH} & = & \frac{m_{\phi}^2 }{48 \pi ^2}\left[\ln \left(\frac{m_{\phi}^2}{\lambda ^2}\right)-2\right],\\
h^{\phi}_{HT} & = &\frac{1}{216 \pi ^2} [
-14 \delta ^3+(6 \delta ^3-9 \delta  m_{\phi}^2 ) \ln \left(\frac{m_{\phi}^2}{\lambda ^2}\right)+6 \left(m_{\phi}^2-\delta ^2\right)^{3/2} \arccos\left(\frac{\delta }{m_{\phi}}\right)\nonumber\\
&&-6 \left(m_{\phi}^2-\delta ^2\right)^{3/2}
\arccos\left(-\frac{\delta }{m_{\phi}}\right)+18 \delta
m_{\phi}^2], \\
m^{\phi}_{H}&=&\frac{1}{16 \pi ^2} [ 2 \delta ^2+\left(m_{\phi}^2-2
\delta ^2\right) \ln \left(\frac{m_{\phi}^2}{\lambda ^2}\right)+4
\delta \sqrt{m_{\phi}^2-\delta ^2} \arccos\left(\frac{\delta
}{m_{\phi}}\right)],\\
n^{\phi}_{H}&=&o^{\phi}_{H}=\frac{m_{\phi}^2 }{16 \pi ^2}\ln \left(\frac{m_{\phi}^2}{\lambda ^2}\right),\\
p^{\phi}_{T}&=&\frac{1}{16 \pi ^2}[ 2 \delta ^2+\left(m_{\phi}^2-2
\delta ^2\right) \ln \left(\frac{m_{\phi}^2}{\lambda ^2}\right)-4
\delta \sqrt{m_{\phi}^2-\delta ^2} \arccos\left(-\frac{\delta
}{m_{\phi}}\right)].
\end{eqnarray}
where $\mud=4\pi f_{\pi}$ is the renormalization scale. The
coefficients $\beta^\phi_{a}$ and $\beta^\phi_{b}$ arise from the
spin-$\frac{3}{2}$ and spin-$\frac{1}{2}$ doubly charmed baryon
intermediate states respectively.
$\gamma^\phi_{c}$,$\gamma^\phi_{d}$,$\gamma^\phi_{e}$,
$\gamma^\phi_{f}$, $\gamma^\phi_{TH}$ and $\gamma^\phi_{w}$ arise
from the corresponding diagrams in Fig.~\ref{fig:allloop}. We
collect their explicit expressions in Tables~\ref{table:abcd},
\ref{table:ef}, \ref{table:gh} in the Appendix \ref{appendix-A}.

With the low energy counter terms and loop contributions
(\ref{eq:muTN2Loop}, \ref{eq:muTN3Loop}), we obtain the transition
magnetic moments,
\begin{equation}
\mu_{TH}=\left\{\mu_{TH}^{(1)}\right\}+\left\{\mu_{TH}^{(2,\rm
loop)}\right\}+\left\{\mu_{TH}^{(3,\rm tree)}+\mu_{TH}^{(3,\rm
loop)}\right\}
\end{equation}
where $\mu_{TH}^{(1)}$ and $\mu_{TH}^{(3,\rm tree)}$ are the
tree-level magnetic moments from
Eqs.~(\ref{Eq:baryon_trans})-(\ref{Eq:MM3}).

\section{NUMERICAL RESULTS AND DISCUSSIONS}\label{Sec6}

\begin{table}
  \centering
\begin{tabular}{c|cc}
\toprule[1pt]\toprule[1pt] Process & $\mathcal{O}(p^{1})$ tree &
$\mathcal{O}(p^{2})$ loop \tabularnewline \midrule[1pt]
$\Xi_{cc}^{*++}\rightarrow\Xi_{cc}^{++}$ &
$(\frac{2}{3}+\frac{\delta}{6M_{T}})(\frac{2}{3}a_{5}+4a_{6})$ &
$(-0.15\tilde{g}_{A}-0.31\tilde{g}_{B})\tilde{g}_{C}$
\tabularnewline

$\Xi_{cc}^{*+}\rightarrow\Xi_{cc}^{+}$ &
$(\frac{2}{3}+\frac{\delta}{6M_{T}})(-\frac{1}{3}a_{5}+4a_{6})$ &
$(0.04\tilde{g}_{A}+0.11\tilde{g}_{B})\tilde{g}_{C}$\tabularnewline

$\Omega_{cc}^{*+}\rightarrow\Omega_{cc}^{+}$ &
$(\frac{2}{3}+\frac{\delta}{6M_{T}})(-\frac{1}{3}a_{5}+4a_{6})$ &
$(0.12\tilde{g}_{A}+0.20\tilde{g}_{B})\tilde{g}_{C}$\tabularnewline
\bottomrule[1pt]\bottomrule[1pt]
\end{tabular}
\caption{The spin-$\frac{3}{2}$ to spin-$\frac{1}{2}$ doubly charmed
baryon transition magnetic moments to the next-to-leading order (in
unit of $\mu_N$).} \label{Magnetic moments}
\end{table}

There are not any experimental information on the spin-$\frac{3}{2}$
to spin-$\frac{1}{2}$ doubly charmed baryon transitions yet. Thus,
we adopt the same strategy as in Ref. \cite{Li3}. We use the quark
model to estimate the leading-order tree level transition magnetic
moments. At the leading order $\mathcal{O}(p^{1})$, as the charge
matrix $Q_{H}$ is not traceless, there are two unknown LECs
$a_{5,6}$ in Eq.~(\ref{Eq:baryon_trans}). Notice the second column
in Table~\ref{Magnetic moments}, the $a_5$ parts are proportional to
the light quark charge within the doubly charmed baryon and the
$a_6$ parts are the same for all three transition processes.

At the quark level, the flavor and spin wave functions of the
spin-$\frac{1}{2}$ and spin-$\frac{3}{2}$ doubly charmed baryons
$\Xi_{ccq}$ and $\Xi_{ccq}^{*}$ read:
\begin{eqnarray}
|\Xi_{ccq};s_3=\frac{1}{2}\rangle&=&\frac{1}{3\sqrt{2}}[2c\uparrow
c\uparrow q\downarrow-c\uparrow c\downarrow q\uparrow-c\downarrow
c\uparrow q\uparrow +2c\uparrow q\downarrow c\uparrow-c\downarrow
q\uparrow c\uparrow\nonumber\\&&-c\downarrow q\downarrow
c\downarrow+2q\downarrow c\uparrow c\uparrow-q\downarrow c\downarrow
c\downarrow-q\uparrow c\downarrow c\uparrow],\label{xiwavefunc}\\
|\Xi_{ccq}^{*};s_3=\frac{1}{2}\rangle&=&\frac{1}{\sqrt{3}}[
c\uparrow c\uparrow q\downarrow+c\uparrow c\downarrow
q\uparrow+c\downarrow c\uparrow q\uparrow],\label{xi*wavefunc}
\end{eqnarray}
where the arrows denote the third-components of the spin $s_3$. $q$
can be $u$,$d$ and $s$ quark. The spin-$\frac{3}{2}$ to
spin-$\frac{1}{2}$ doubly charmed baryon transition magnetic moments
in the quark model are the matrix elements of the following operator
between Eq.~(\ref{xiwavefunc}) and Eq.~(\ref{xi*wavefunc}),
\begin{equation}
\vec{\mu}=\sum_i\mu_i\vec{\sigma}^i, \label{magmomen}
\end{equation}
where $\mu_i$ is the magnetic moment of the quark:
\begin{equation}
\mu_i={e_i\over 2m_i},\quad i=u,d,s,c.
\end{equation}
We adopt the $m_u=m_d=336$ MeV, $m_s=540$ MeV, $m_c=1660$ MeV as the
constituent quark masses and give the results in the second column
in Table~\ref{various orders Magnetic moments}. The light quark
magnetic moments contributes to the LEC $a_{5}$, which is
proportional to the light quark charge. The heavy quark magnetic
moments contributes to the LEC $a_{6}$, which are the same for all
three transitions.

\begin{table}
  \centering
\begin{tabular}{c|ccc}
\toprule[1pt]\toprule[1pt] Process& $\mathcal{O}(p^{1})$ &
$\mathcal{O}(p^{2})$ loop & $\mathcal{O}(p^{2})$Total\tabularnewline
\midrule[1pt] $\Xi_{cc}^{*++}\rightarrow\Xi_{cc}^{++}$ &
$\frac{1}{3\sqrt{2}}(4\mu_{c}-4\mu_{u})=-1.40$ &
$(-0.15\tilde{g}_{A}-0.31\tilde{g}_{B})\tilde{g}_{C}=-0.95$ &
-2.35\tabularnewline

$\Xi_{cc}^{*+}\rightarrow\Xi_{cc}^{+}$ &
$\frac{1}{3\sqrt{2}}(4\mu_{c}-4\mu_{d})=1.23$ &
$(0.04\tilde{g}_{A}+0.11\tilde{g}_{B})\tilde{g}_{C}=0.32$ &
1.55\tabularnewline

$\Omega_{cc}^{*+}\rightarrow\Omega_{cc}^{+}$ &
$\frac{1}{3\sqrt{2}}(4\mu_{c}-4\mu_{s})=0.90$ &
$(0.12\tilde{g}_{A}+0.20\tilde{g}_{B})\tilde{g}_{C}=0.64$ &
1.54\tabularnewline \bottomrule[1pt]\bottomrule[1pt]
\end{tabular}
\caption{The spin-$\frac{3}{2}$ to spin-$\frac{1}{2}$ doubly charmed
baryon transition magnetic moments when $\delta=0.1$GeV (in unit of
$\mu_{N}$).} \label{various orders Magnetic moments}
\end{table}

Up to $\mathcal{O}(p^{2})$, we need include both the leading
tree-level magnetic moments and the $\mathcal{O}(p^{2})$ loop
corrections. We use the quark model to estimate the leading-order
tree level transition magnetic moments. Thus, there exist only three
LECs: $\tilde{g}_{A}$, $\tilde{g}_{B}$ and $\tilde{g}_{C}$ at this
order. $\tilde{g}_{A}=-\frac{2}{5}g_{A}=-0.50$ has been estimated in
Ref. \cite{Li3}. After similar calculations, one obtains the $\phi T
T$ coupling $\tilde{g}_{B}=-\frac{6}{5}g_{A}=-1.51$ and $\phi T H$
coupling $\tilde{g}_{C}=-\frac{4\sqrt{3}}{5}g_{A}=-1.75$. With these
LECs, we obtain the numerical results of the $\mathcal{O}(p^{2})$
spin-$\frac{3}{2}$ to spin-$\frac{1}{2}$ doubly charmed baryon
transition magnetic moments in Table~\ref{various orders Magnetic
moments}. As the E2 transtion moments are much smaller than M1
transtion moments, using Eq. (\ref{eq:decaywidth}) we can also
calculate the decay width of the spin-$\frac{3}{2}$ to
spin-$\frac{1}{2}$ doubly charmed baryon transitions in
Table~\ref{decay width}.

Up to $\mathcal{O}(p^{3})$, there are ten unknown LECs: $a_{1-6}$,
$h_1$, $d_{1,2,3}$. It is impossible to to fix all these LECs with
the present experimental data. We present our analytical results in
Eqs. (\ref{eq:muTN2Loop})-(\ref{eq:muTN3Loop}) and
Table~\ref{Magnetic moments}.

We also calculate the spin-${3\over 2}$ to spin-${1\over2}$
transition magnetic moments for the doubly bottomed baryons and
charmed bottomed baryons. The results are given in
Table~\ref{Morment bb}. For the charmed bottomed baryons, we use the
subscript $\{bc\}$ and $[bc]$ to label the systems with symmetric
and antisymmetry spin wave functions in the heavy quark sector,
respectively. It is very interesting to note that the light quark
does not contribute to the $(\{bc\}q)^*\rightarrow([bc]q)$
transition magnetic moments at the tree-level in the quark model.
Meanwhile, for these processes, the $\mathcal{O}(p^2)$ loop
contribution also vanishes for the vanishing axial coupling
constants. In the quark model, the pions only couple to the light
quark sector of the heavy baryons. However, the heavy quark spin
wave functions of the $(\{bc\}q)^*$ and $([bc]q)$ systems are
orthogonal to each other. Thus, the pionic couplings vanish.

\begin{table}
  \centering
\begin{tabular}{c|c}
\toprule[1pt]\toprule[1pt] Process&  Decay width
$\Gamma$/\rm{keV}\tabularnewline \midrule[1pt]
$\Xi_{cc}^{*++}\rightarrow\Xi_{cc}^{++}$ & $22.0$\tabularnewline

$\Xi_{cc}^{*+}\rightarrow\Xi_{cc}^{+}$  & $9.57$\tabularnewline

$\Omega_{cc}^{*+}\rightarrow\Omega_{cc}^{+}$ & $9.45$\tabularnewline
\bottomrule[1pt]\bottomrule[1pt]
\end{tabular}
\caption{The spin-$\frac{3}{2}$ to spin-$\frac{1}{2}$ doubly charmed
baryon decay width when $|q|=0.1$GeV (in unit of keV).} \label{decay
width}
\end{table}

\begin{table}
  \centering
\begin{tabular}{c|cccc}
\toprule[1pt]\toprule[1pt]
Process & ${\cal O}(p^{1})$ & ${\cal O}(p^{2})$ loop & ${\cal O}(p^{2})$ Total & Decay width $\Gamma$/keV\tabularnewline
\midrule[1pt]
$\Xi_{bb}^{*0}\rightarrow\Xi_{bb}^{0}$ & $\frac{2\sqrt{2}}{3}\left(\mu_{b}-\mu_{u}\right)=-1.82$ & -0.95 & -2.77 & 31.1\tabularnewline

$\Xi_{bb}^{*-}\rightarrow\Xi_{bb}^{-}$ & $\frac{2\sqrt{2}}{3}\left(\mu_{b}-\mu_{d}\right)=0.81$ & 0.32 & 1.13 & 5.17\tabularnewline

$\Omega_{bb}^{*-}\rightarrow\Omega_{bb}^{-}$ & $\frac{2\sqrt{2}}{3}\left(\mu_{b}-\mu_{s}\right)=0.48$ & 0.64 & 1.12 & 5.08
\tabularnewline

$\Xi_{\{bc\}}^{*+}\rightarrow\Xi_{\{bc\}}^{+}$ & $\frac{\sqrt{2}}{3}\left(\mu_{c}+\mu_{b}-2\mu_{u}\right)=-1.61$ & -0.95 & -2.56 & 26.2\tabularnewline

$\Xi_{\{bc\}}^{*0}\rightarrow\Xi_{\{bc\}}^{0}$ & $\frac{\sqrt{2}}{3}\left(\mu_{c}+\mu_{b}-2\mu_{d}\right)=1.02$ & 0.32 & 1.34 & 7.19\tabularnewline

$\Omega_{\{bc\}}^{*0}\rightarrow\Omega_{\{bc\}}^{0}$ & $\frac{\sqrt{2}}{3}\left(\mu_{c}+\mu_{b}-2\mu_{s}\right)=0.69$ & 0.64 & 1.33 & 7.08
\tabularnewline

$\Xi_{\{bc\}}^{*+}\rightarrow\Xi_{[bc]}^{+}$ & $\sqrt{\frac{2}{3}}\left(\mu_{b}-\mu_{c}\right)=-0.36$ & 0 & -0.36 & 0.52\tabularnewline

$\Xi_{\{bc\}}^{*0}\rightarrow\Xi_{[bc]}^{0}$ & $\sqrt{\frac{2}{3}}\left(\mu_{b}-\mu_{c}\right)=-0.36$ & 0 & -0.36 & 0.52\tabularnewline

$\Omega_{\{bc\}}^{*0}\rightarrow\Omega_{[bc]}^{0}$ & $\sqrt{\frac{2}{3}}\left(\mu_{b}-\mu_{c}\right)=-0.36$ & 0 & -0.36 & 0.52\tabularnewline \bottomrule[1pt]\bottomrule[1pt]
\end{tabular}
\caption{The spin-$\frac{3}{2}$ to spin-$\frac{1}{2}$ doubly
bottomed baryon transition magnetic moments when $\delta=0.1$GeV (in
unit of $\mu_{N}$). For the charmed bottomed baryons, we use the
subscripts $\{bc\}$ and $[bc]$ to label the systems with symmetric
and antisymmetry spin wave functions in the heavy quark sector,
respectively. } \label{Morment bb}
\end{table}

\section{Conclusions}\label{Sec7}

\begin{table}
  \centering
\begin{tabular}{c|ccccc}
\toprule[1pt]\toprule[1pt] Transitions&  ENRQM~\cite{Dhir:2013nka} &
$\chi$CQM~\cite{Sharma:2010vv} & MIT bag
model~\cite{Bernotas:2013eia}&NRQM&This work\tabularnewline
\midrule[1pt] $\Xi_{cc}^{*++}\rightarrow\Xi_{cc}^{++}$ &
1.35&1.33&-0.787&-1.40&-2.35\tabularnewline

$\Xi_{cc}^{*+}\rightarrow\Xi_{cc}^{+}$  &
1.06&-1.41&0.945&1.23&1.55\tabularnewline

$\Omega_{cc}^{*+}\rightarrow\Omega_{cc}^{+}$ &
0.88&-0.89&0.789&0.90&1.54\tabularnewline
\bottomrule[1pt]\bottomrule[1pt]
\end{tabular}
\caption{Comparison of the spin-$\frac{3}{2}$ to spin-$\frac{1}{2}$
doubly charmed baryon transition magnetic moments in literature
including nonrelativistic quark model with effective quark mass
(SCR)~\cite{Dhir:2013nka}, SU(4) chiral constituent quark model
($\chi$CQM)~\cite{Sharma:2010vv}, MIT bag
model~\cite{Bernotas:2013eia} and nonrelativistic quark model(NRQM)
(in unit of $\mu_{N}$).} \label{compare}
\end{table}

The recently observation of $\Xi_{cc}^{++}$ has aroused tremendous
attention to the doubly charmed baryons. The chiral dynamics of the
doubly charmed baryons is simpler than the nucleon, which allows one
to study the chiral dynamics of the light quarks more directly.
Moreover, the spin-$\frac{3}{2}$ to spin-$\frac{1}{2}$ doubly
charmed baryon transition electromagnetic properties probe the inner
structure and possible deformation of both the spin-$\frac{1}{2}$
and spin-$\frac{3}{2}$ doubly charmed baryons.

In this paper, we have systematically calculated the chiral
corrections to the spin-$\frac{3}{2}$ to spin-$\frac{1}{2}$ doubly
charmed baryon transition magnetic moments up to the
next-to-next-to-leading order in the framework of the heavy baryon
chiral perturbation theory. In principle, the low energy constants
at the tree-level should be extracted through fitting to the
experimental data. However, the current experimental information is
very poor. With the help of quark model, we have estimated the LECs
such as the leading-order axial coupling and tree level transition
magnetic moments. Because of lack of enough experimental inputs, we
only present the numerical results up to next-to-leading order:
$\mu_{\Xi_{cc}^{*++}\rightarrow\Xi_{cc}^{++}}=-2.35\mu_{N}$,
$\mu_{\Xi_{cc}^{*+}\rightarrow\Xi_{cc}^{+}}=1.55\mu_{N}$,
$\mu_{\Omega_{cc}^{*+}\rightarrow\Omega_{cc}^{+}}=1.54\mu_{N}$. In
Table~\ref{compare}, we compare our results in HBChPT with those
from other approaches such as nonrelativistic quark model with
effective quark mass (ENRQM)~\cite{Dhir:2013nka}, SU(4) chiral
constituent quark model ($\chi$CQM)~\cite{Sharma:2010vv}, MIT bag
model~\cite{Bernotas:2013eia} and nonrelativistic quark model(NRQM).

We also calculate the decay width of the doubly charmed baryon
transitions: $\Gamma_{\Xi_{cc}^{*++}\rightarrow\Xi_{cc}^{++}}=22.0$
keV, $\Gamma_{\Xi_{cc}^{*+}\rightarrow\Xi_{cc}^{+}}=9.57$ keV,
$\Gamma_{\Omega_{cc}^{*+}\rightarrow\Omega_{cc}^{+}}=9.45$ keV. As
we use the quark model to estimate the leading-order tree level
transition magnetic moments, the final transition magnetic moments
and decay width may be enhanced to some extent.

We hope our results may be useful for future experimental
measurement of the transition magnetic moments. Moreover, the
analytical expressions derived in this work may be useful to the
possible chiral extrapolation of the lattice simulations of the
double charmed baryon transition electromagnetic properties in the
coming future.

\section*{ACKNOWLEDGMENTS}

H. S. Li is very grateful to X. L. Chen and W. Z. Deng for very
helpful discussions. This project is supported by the National
Natural Science Foundation of China under Grants 11575008,
11621131001 and 973 program. This work is also supported by the
Fundamental Research Funds for the Central Universities of Lanzhou
University under Grants 223000--862637.

\begin{appendix}

\section{COEFFICIENTS OF THE LOOP CORRECTIONS} \label{appendix-A}

We collect the explicit formulae for the chiral expansion of the
spin-$\frac{3}{2}$ to spin-$\frac{1}{2}$ doubly charmed baryon
transform magnetic moments in this appendix.

\begin{table}
  \centering
\begin{tabular}{c|cccccccc}
\toprule[1pt]\toprule[1pt] Process & $\beta_{a}^{(\pi)}$ &
$\beta_{a}^{(K)}$ & $\beta_{b}^{(\pi)}$ & $\beta_{b}^{(K)}$ &
$\gamma_{c}^{(\pi)}$ & $\gamma_{c}^{(K)}$ & $\gamma_{d}^{(\pi)}$ &
$\gamma_{d}^{(K)}$\tabularnewline \midrule[1pt]
$\Xi_{cc}^{*++}\rightarrow\Xi_{cc}^{++}\gamma$ & $2$ & $2$ & $2$ &
$2$ & $-4a_{5}$ & $-4a_{5}$ & $2h_{1}$ & $2h_{1}$\tabularnewline

$\Xi_{cc}^{*+}\rightarrow\Xi_{cc}^{+}\gamma$ & $-2$ & 0 & $-2$ & 0 &
$4a_{5}$ & 0 & $-2h_{1}$ & 0\tabularnewline

$\Omega_{cc}^{*+}\rightarrow\Omega_{cc}^{+}\gamma$ & 0 & $-2$ & 0 &
$-2$ & 0 & $4a_{5}$ & 0 & $2h_{1}$\tabularnewline
\bottomrule[1pt]\bottomrule[1pt]
\end{tabular}
\caption{The coefficients of the loop corrections to the
 doubly charmed baryon magnetic moments from Figs.
\ref{fig:allloop}(a), \ref{fig:allloop}(b), \ref{fig:allloop}(c) and
\ref{fig:allloop}(d).} \label{table:abcd}
\end{table}

\begin{table}
  \centering
\begin{tabular}{c|cccccc}
\toprule[1pt]\toprule[1pt] Process & $\gamma_{e}^{(\pi)}$ &
$\gamma_{e}^{(K)}$ & $\gamma_{e}^{(\eta)}$ & $\gamma_{f}^{(\pi)}$ &
$\gamma_{f}^{(K)}$ & $\gamma_{f}^{(\eta)}$\tabularnewline
\midrule[1pt] $\Xi_{cc}^{*++}\rightarrow\Xi_{cc}^{++}\gamma$ &
$12a_{4}$ & $-\frac{2}{3}a_{3}+8a_{4}$ & $\frac{2}{9}(a_{3}+6a_{4})$
& $12a_{2}$ & $-\frac{2}{3}a_{1}+8a_{2}$ &
$\frac{2}{9}(a_{1}+6a_{2})$\tabularnewline

$\Xi_{cc}^{*+}\rightarrow\Xi_{cc}^{+}\gamma$ & $a_{3}+12a_{4}$ &
$-\frac{2}{3}a_{3}+8a_{4}$ & $\frac{1}{9}(-a_{3}+12a_{4})$ &
$a_{1}+12a_{2}$ & $-\frac{2}{3}a_{1}+8a_{2}$ &
$\frac{1}{9}(-a_{1}+12a_{2})$\tabularnewline

$\Omega_{cc}^{*+}\rightarrow\Omega_{cc}^{+}\gamma$ & 0 &
$\frac{2}{3}(a_{3}+24a_{4})$ & $-\frac{4}{9}(a_{3}-12a_{4})$ & 0 &
$\frac{2}{3}(a_{1}+24a_{2})$ &
$-\frac{4}{9}(a_{1}-12a_{2})$\tabularnewline
\bottomrule[1pt]\bottomrule[1pt]
\end{tabular}
\caption{The coefficients of the loop corrections to the transform
magnetic moments from Figs. \ref{fig:allloop}(e) and
\ref{fig:allloop}(f).} \label{table:ef}
\end{table}

\begin{table}
  \centering
\begin{tabular}{c|cccccc}
\toprule[1pt]\toprule[1pt] Process & $\gamma_{TH}^{(\pi)}$ &
$\gamma_{TH}^{(K)}$ & $\gamma_{TH}^{(\eta)}$ & $\gamma_{w}^{(\pi)}$
& $\gamma_{w}^{(K)}$ & $\gamma_{w}^{(\eta)}$\tabularnewline
\midrule[1pt] $\Xi_{cc}^{*++}\rightarrow\Xi_{cc}^{++}\gamma$ &
$12a_{6}$ & $-\frac{2}{3}a_{5}+8a_{6}$ & $\frac{2}{9}(a_{5}+6a_{6})$
& $2(a_{5}+6a_{6})$ & $\frac{4}{3}(a_{5}+6a_{6})$ &
$\frac{2}{9}(a_{5}+6a_{6})$\tabularnewline

$\Xi_{cc}^{*+}\rightarrow\Xi_{cc}^{+}\gamma$ & $a_{5}+12a_{6}$ &
$-\frac{2}{3}a_{5}+8a_{6}$ & $\frac{1}{9}(-a_{5}+12a_{6})$ &
$-a_{5}+12a_{6}$ & $-\frac{2}{3}a_{5}+8a_{6}$ &
$-\frac{1}{9}a_{5}+\frac{4}{3}a_{6}$\tabularnewline

$\Omega_{cc}^{*+}\rightarrow\Omega_{cc}^{+}\gamma$ & 0 &
$\frac{2}{3}(a_{5}+24a_{6})$ & $-\frac{4}{9}(a_{5}-12a_{6})$ & 0 &
$-\frac{4}{3}a_{5}+16a_{6}$ &
$-\frac{4}{9}(a_{5}-12a_{6})$\tabularnewline
\bottomrule[1pt]\bottomrule[1pt]
\end{tabular}
\caption{The coefficients of the loop corrections to the
 doubly charmed baryon magnetic moments from Figs.
\ref{fig:allloop}(g-h) and \ref{fig:allloop}(m-p).} \label{table:gh}
\end{table}

\end{appendix}

\vfil \thispagestyle{empty}

\end{document}